\documentstyle[12pt,epsf]{article}
\begin{document}

\title{Mass Hierarchy and Mixing as Results of Simultaneous Dynamical
Breaking of Chiral and Flavor Symmetries}
\author{I.T. Dyatlov \\ Petersburg Nuclear Physics Institute\\
Gatchina, St.Petersburg 188350, Russia, dyatlov@thd.pnpi.spb.ru}
\date{}
\maketitle

\begin{abstract}
If the hierarchy of quark masses comes out of simultaneous dynamical
chiral and flavor symmetry breakings, it is inevitably accompanied by
the flavor mixing. Therefore, these phenomena, independent in the
standard model and in many of its generalizations, become related since
they appear out of the same dynamics. The features of that high energy
dynamics may be investigated via accordance with observed mass
hierarchy and mixing.
\end{abstract}

The $t$-condensate approach to chiral (and weak) symmetry breaking
seems enable reasonably to reproduce the standard model (SM) properties
concerning masses of the heavy quark, gauge bosons and the Higgs scalar
\cite{1}. This fact could evidence that the physical mechanism based on
the Nambu's idea \cite{2} is close to a true one, that is really
responsible for the origin of SM quark masses.

Limitations of the method consist in leaving aside other quark masses
(besides the most heavy $t$-one) and neglecting mixing between quark
generations. To incorporate the both phenomena new  hypotheses and new
independent dynamical mechanisms are required. This defect results from
the principal idea about a special strong coupling acting exclusively
within the $t$-generation. This idea is in contrast with total flavor
symmetry of all observed interactions.

Therefore, it seems that a more attractive and promising picture would
evolve if the most heavy mass can appear via simultaneous chiral and
flavor symmetry breakings in a system of $n$ initially massless and
symmetric  quark flavors ($n=3,4,\ldots$). Then a would be ground
state with one massive $(m_1\neq0)$ and $n-1$ massless quarks
$(m_2=m_3=\cdots=0)$ turns out the very state to develop mass hierarchy
and mixing. A small perturbative violation of the initial symmetries
easily produces the necessary development. The phenomena automatically
appear together due to a common dynamical mechanism. Properties of the
 involved perturbative dynamics may be then extracted out of the
coincidence with the observed picture for quark mass spectra and,
especially, for the weak mixing.

Thus, we are interested to find out appropriate conditions for a phase
transition of the type ($L,R$ are chiralities):
\begin{equation}
U_L(n)\times U_R(n)\ \to\ U_L(n-1)U_R(n-1)U(1)\ ,
\end{equation}
in a system of $n$ identical massless quark flavors. As the simplest
possible dynamical model let us consider the four-\-fermion interaction
\begin{equation}
V_{int}=\ \lambda \sum_{i,k;c,c'}\ (\bar\psi^c_{iL}\psi^c_{kR})
(\bar\psi^{c'}_{kR}\psi^{c'}_{iL})\ ,
\end{equation}
$i,k=1,2,\ldots,n$ are $L,R$ flavors; $cc'=1,2,\ldots,N_c$ are colors.
Our system will be one of the Nambu-\-Yona-\-Lasinio (NYL) models
\cite{3}. The last ones have been often used before within
investigations of the $t$-condensate mechanics \cite{1}--\cite{4}. The
solution of the model is usually looked for as an expansion in
$N^{-1}_c$, $N_c\gg1$, with a dimensionless coupling
\begin{equation}
\beta\ =\ \frac{\lambda N_c M^2}{8\pi^2}\ ,
\end{equation}
taken as a constant at $N_c\to\infty$.

The factor $M$ is a momentum cut-off of the unrenormalizable model. Let
us remind that principal features of chiral violation: initiation of
masses and emergence of goldstone bosons, --- insignificantly depend on
a detailed behaviour near the cut-\-off region $M$. However, those
$N^{-1}_c$ orders $(M_c^0,N_c^{-1})$, where consistent treatment is
possible in pure NYL models \cite{5}, are inadequate to achieve our
 new aims.

Several physical reasons dictate the choice of Eq.(2) as a reasonable
trial device.

1. Eq.(2) is an $U_L(n)\times U_R(n)$ symmetric form. Here the flavor
groups coincide with chiral ones. Only in such a system simultaneous
breaking can occur.

2. The Fierz transformation carries (2) into a combination of (color)
vector-\-vector $L\times R$ couplings conserving flavors. Therefore (2)
may be considered as a low-\-energy local effective limit
corresponding to a high-\-energy theory where all interactions are
generated exclusively by vector (pseudovector) fields. Indeed, any
number of vector exchanges between massless chiral fermions generates
only a vector-\-vector coupling in the local limit. As it is usually
supposed, the gauge theories are favorable candidates in dynamical
explaining of SM properties. Therefore the potential (2) may present a
rather suitable example for a low-\-energy limit of such a theory.

3. In addition (although it is not so significant), among all flavor
conserving local couplings only the colorless, scalar form (2)
contributes to the equation for fermion masses (the "gap
equation"\cite{3}) in the leading $N^0_c$ approximation.

Now let us construct NYL gap equations. In the symmetric problem (1)
the matrix of quark propagators can be taken in a diagonal form
\begin{equation}
G^{-1}_{ii}(p)\ = \ \Sigma_{ii}(p^2)-\hat pZ_{ii}(p^2)\ .
\end{equation}
$\Sigma$ and $Z$ are quantities to be determined by the equations. In
the leading (Fig.1) and next-\-after-\-leading (Fig.2) $N^{-1}_c$
approximations\footnote{In all approximations, besides the leading one,
equations for a self-\-energy $\Sigma$ and for a mass matrix are
different.} gap-\-equations show possible existence of various phase
transitions
\begin{equation}
U_L(n)\times U_R(n)\ \to\ U_L(n-n')\times U_R(n-n')U_V(n')\ ,
\end{equation}
$n'\le n$. Any such a solution contains $n'$ massive flavors:
\begin{equation}
m_1\ =\ m_2\ =\ \cdots\ =\ m_{n'}\ \neq\ 0\ ,
\end{equation}
and $n-n'$ massless ones
\begin{equation}
m_{n'+1}\ =\ \cdots\ =\ m_n\ =\ 0\ .
\end{equation}

In the lowest approximation the gap-equations take the well-\-known
\cite{1}--\cite{4} form
\begin{equation}
m_i\ =\ m_i\ \frac\beta{\pi^2M^2}\int \frac{d^4p\,f(M^2-|p|^2)}{m^2_i
+p^2}\ ,
\end{equation}
$M^2$ is the momentum cut-off, $f$ is the cut-off function. We
transformed the integral (8) into the Euclidean form. The $\beta$
parameter is the usual for NYL the dimensionless coupling constant (3).

We shall confirm now the presence of the phase transitions (5),
constructing compound bosons of the model. A $q\bar q$ scattering
amplitude (Fig.3)
\begin{equation}
F(p_i)\ =\ \frac\beta{N_cM^2}\ B(q)
\end{equation}
depends on a single variable $q$ in the lowest $N^{-1}_c$
approximation. Summing loops, it is easy to show that chirality
conserving $(B_+)$ and changing $(B_-)$ parts are
\begin{equation}
B_{\pm}\ =\ \frac12\left[ \frac1{1-A_1+A_2} \pm \frac1{1-A_1-A_2}
\right]\ ,
\end{equation}
if we have written the one-loop contribution as
\begin{equation}
A_{\alpha_1\alpha_2}\ =\ A_1\delta_{\alpha_1\alpha_2}
+A_2\delta_{\alpha_1-\alpha_2}\ , \qquad \alpha_i=(L,R) =\pm1\ .
\end{equation}

In the denominators (10) let us remove the part reproducing the gap
equation(8). We have
\begin{eqnarray}
1-A_1\pm A_2&=&1 -\frac\beta{M^2}\int\frac{d^4p}{\pi^2i}
\frac{f(M^2-|p|^2)}{m^2_1-p^2}+\frac\beta{2M^2}\int\frac{d^4p}{\pi^2i}
\left[\frac1{m^2_1-p^2}-\frac1{m^2_2-(p-q)^2}\right]f \nonumber \\
&& +\ \;\; \frac\beta{2M^2}\left[(m_1\pm m_2)^2-q^2\right]\ I(m,q)\ .
\end{eqnarray}
We consider the scattering amplitude for quarks with $m_1$ and $m_2$
masses. The function $I(m,q)$ is
\begin{equation}
I(m,q)\ =\ \int \frac{d^4p}{\pi^2i}\
\frac1{(m^2_1-p^2)[m^2_2-(p-q)^2]}f\ .
\end{equation}
The first two terms (12)
vanish, when $m_1$ is a root of the equation(8) and this result does
not depend on the cut-\-off choice.  All other terms become
\begin{equation}
\frac\beta{2M^2}\int\frac{d^4p}{\pi^2i}\ \frac{(2pq-q^2)+m^2_2-m^2_1
+(m_1\pm m_2)^2-q^2}{(m^2_1-p^2)[m^2_2-(p-q)^2]}\ .
\end{equation}
It is evident from (14) that: \\
for $m_1=m_2$:
\begin{eqnarray}
1-A_1+A_2&=&\frac\beta{2M^2}\left[\frac12(-q^2)+(4m^2_1-q^2)I
(m_1,q)\right]\ , \\
1-A_1-A_2 &=& \frac\beta{2M^2}\left[\frac12(-q^2)+(-q^2)I
(m_1,q)\right]\ ;
\end{eqnarray}
and for $m_1\neq0$, $m_2=0$:
\begin{equation}
1-A_1\pm A_2\ =\ \frac\beta{2M^2}\left[\frac12(-q^2)+(-q^2)I
(m_1,q)\right]\ .
\end{equation}

Combining all massive quarks between themselves and massive with
massless ones we obtain the number of compound massless bosons in the
model: $n'(4n-3n')$. It is just the number one can obtain calculating
Goldstone modes associated with the transition (5):
\begin{equation}
2n^2-2(n-n')^2-n'^2\ =\ n'(4n-3n')\ .
\end{equation}
Thus, gap-equations of the NYL model evidence possibility of the
transitions (5). They can be traced in both consistent approximations
of the NYL models $(N^0_c,N^{-1}_c)$. Then, the remaining question is:
what a state will be vacuum one (i.e. the lowest energy state)~?

The leading $N^0_c$ approximation determines as the vacuum state the
flavor symmetric solution $n'=n$. This can be proved by constructing a
corresponding effective potential. In the next $N^{-1}_c$ approximation
it was stated that the $n'=1$ solution may be the vacuum state at some
conditions. It was confirmed in \cite{6} by lattice calculations. But
\cite{6} did not take into account significant part (17) of Goldstone
modes (when massive and massless quarks are combined). Nevertheless,
next $N^{-1}_c$ approximations  show that the dependence of $n'$
inevitably appears in the NYL gap equations. Prompting by these
considerations and possible essential contributions from higher $N_c$
approximations, a general form of the gap-\-equation could be imagined
as
\begin{equation}
1-\beta^{-1}=\frac{m^2_1}{M^2}\ln\frac{M^2}{m^2_1}+\frac{n'}{N_c}\,
f_1\left(N_c,\frac{m^2_1}{M^2}\right)+\frac1{N_c}\,f_2\left(N_c,
\frac{m^2_1}{M^2}\right)\ .
\end{equation}
Here $f_1$ and $f_2$ are some functions which cannot be calculated in
NYL models for high $N^{-1}_c$ approximations. They essentially depend
on the region near the cut-\-off $M$. The term $f_1$ is crucial. If
$f_1>0$ and weakly (less than $m^2_1/M^2)$ depends on $m^2_1/M^2$, a
region $(\beta)$ appears where only the $n'=1$ solution becomes
possible.

Therefore, let us assume that in a flavor symmetric, massless system
the vacuum state with one massive and $n-1$ massless quarks becomes
possible. That state will realize simultaneous breaking of flavor and
chiral symmetries. Then, a single  mechanism relates mass spectra with
mixing properties. Such a relation permits unambiguously to propose and
investigate those interactions at high energies which could be
responsible for the observed mass and CKM matrix properties. In the
 considered approximation of the local low-\-energy interaction (2) we
shall propose and investigate necessary properties of the coupling
constant $\lambda(\beta)$.

In order to insert other masses into a symmetric system, it is
necessary to resolve degeneracy of $(n-1)$ massless quarks. That means
that flavor symmetries must be violated, coupling constants should
depend on flavor indices
\begin{equation}
\lambda\ \to\ \lambda_{ii',kk'}\ , \qquad \beta\ \to\ \beta_{ii',kk'}\
,
\end{equation}
 $ii'$ are $L$ flavors, $k,k'$ are $R$ flavors. But certainly, the
complete absence of flavor symmetries is not a situation, which will be
interesting for our aims. The flavor violation must be small in
comparison with the symmetric (flavor independent) and strong coupling
part $\beta$. Moreover, the only interesting situation seems to be that
one, when flavor numbers remain conserved in (20). There will be the
case when mixing can also appear spontaneously as a result of the same
phase transition and will not be inserted via interaction constants.
Therefore, we choose
\begin{equation}
\beta_{ii',kk'}\ =\ \beta_{ik}\delta_{ii'}\delta_{kk'}\ , \qquad
\beta_{ik}=\ \beta+\delta\beta_{ik}\ , \qquad |\delta\beta_{ik}|\
\ll\ \beta\ .
\end{equation}
The flavor dependent part $\delta\beta_{ik}$ could be perturbative one.

Furthermore, the mass hierarchy property even requires the mass matrix
$M_{ik}$ to be represented as some perturbative series in interactions
distinguishing flavor indices. The last statement is known for years as
"the radiative mechanism" proposed for hierarchy explanations \cite{7}.
If the 4-\-fermion form (2) is a local low-\-energy limit for some high
energy vector (pseudovector) theory, the perturbative and flavor
dependent part of $\beta_{ik}$ must have a form:
\begin{equation}
\beta_{ik}=\ \beta+\delta\beta_{ik}=\ \beta+\sum_{n,m>1}
a_{nm}g_{iL}^ng^m_{kR}\ .
\end{equation}
And $a_{nm}$ coefficients are assumed as decreasing with $n,m$
numbers, similarly to coefficients of any perturbation theory in
$g_L,g_R$ $(\sim c_n(4\pi)^{-n-m}$, $c_n\sim1)$. The formula
(22) becomes evident if one considers a local limit of any number of
vector exchanges between massless $L$ and $R$ fermions $(g_L\neq
g_R\sim1$, Fig.4). Considered vector interactions conserve $R$ and $L$
flavors.

The next step is the sequential perturbative solution \cite{7} of the
gap equation with $\beta_{ik}$, (22) at conditions:
\begin{equation}
|\delta\beta_{ik}|\ \ll\ \beta\ , \qquad m_1\gg m_2\gg m_3\ \ldots\ .
\end{equation}
It is not a simple task for a non-linear equation. But it can be seen
that, when $\beta_{ik}$ is determined by Eq.(22), for the mass matrix
$M_{ik}$ one obtains the solution
\begin{equation}
M_{ik}\ =\ \frac{m_1}n\ R_{ik}+\sum_{n,m\ge0}\ b_{nm}g^n_{iL}g^m_{kR}
\end{equation}
in any order of $N^{-1}_c$ approximations.  The $R$ matrix elements
are: $R_{ik}=1$, for all $i,k$ indices ("democratic matrix", Y.Koide,
H.Fritzsch). Coefficients $b_{nm}$ have the perturbative properties
similar to that ones of $a_{nm}$. They can depend on $g_{iL},g_{kR}$
via transpositionly symmetric combinations (as $\Sigma^n_{i=1}g_{iL}$,
$\Sigma^n_{i=1}g^2_{iL}\ldots$).

The flavor symmetric zero order approximation $\delta\beta_{ik}=0$ for 
the $M_{ik}$ matrix (when only one eigenvalue $m_1$ differs from zero) 
must be taken here in the most general form $(m_1/n)R_{ik}$, since the 
flavor basis has been already fixed by Eq.(21): all interactions are 
diagonal in flavor indices ($L$ and $R$). Therefore, mixing becomes 
inevitable in the situation considered inspite of the flavor conserving 
interactions.  This seems to be an attractive property of the 
simultaneous flavor-\-chirality breaking.

Diagonalization of mass matrices for up and down quark families by
means of unitary matrices $U_L$ and $U_R$:
\begin{equation}
\Sigma^{(u,d(}_{ik}\ =\ U_L^{+(u,d)}\Sigma^{(u,d)}_{diag}U_R^{(u,d)}\ ,
\end{equation}
and calculation of the  weak mixing matrix $V_{CKM}$
\begin{equation}
V_{CKM}\ =\ U^{(d)+}_LU^{(u)}_L
\end{equation}
may be fulfilled for the general form (24) \cite{9}. One obtains the
hierarchy spectrum of masses. It is the direct consequence of the
perturbative structure (22) of $M_{ik}(u)$. An impressive result arises
for the mixing matrix: it automatically acquires a form qualitatively
similar to the observed one (for any coefficients $a_{nm}$ with a
perturbative pattern) when and only when $L$ constants $g^{(u,d)}_{iL}$
are independent of $u,d$ indices:
\begin{equation}
g^{(u)}_{iL}\ =\ g^{(d)}_{iL}\ \equiv\ g_{iL}\ , \qquad g^{(u)}_{kR}\
\neq\ g^{(d)}_{kR}\ .
\end{equation}
The important point is the following. If at high energies one has
some flavor dependent (but neutral, conserving flavor) interaction of
the chiral type $$ \left(\sim
g^{(u,d)}_i\bar\psi^{(u,d)}_{iL}\gamma_\mu\psi^{(u,d)}_{iL} Z'_\mu(x)\
, \qquad \sim g^{(u,d)}_{kR}\bar\psi_{kR}\gamma_\mu\psi_{kR}Z'_\mu(x)
\right)\ ,  $$
the conditions (27) are not new, additional ones, but
those of the $SU_L(2)$ weak symmetry. When $g^{(u)}_{iL}\neq
g^{(d)}_{iL}$ a direct, explicit violation of the weak isospin would be
inserted.

The statement "qualitatively similar" means the following.
\begin{enumerate}
\item The diagonal coefficients $V_{us},V_{cs},V_{tb}$ differ from
unity by the second orders of $g_{iL},g_{kR}$ perturbations.
\item $V_{us}(V_{cd})$ contains terms proportional to the first order
of the perturbations (for the numbers of generations $n=3,4$. Why it is
reasonable to consider $n=4$ --- see below).
\begin{equation}
V_{us}\sim\frac{\langle2nd\rangle}{\langle1st\rangle}
+\langle1st\rangle\ ,\quad n=3\ ;\quad V_{us}\sim\frac{\langle3rd
\rangle+\langle1st\rangle\langle2nd\rangle+\langle 1st\rangle^3}{
\langle1st\rangle^2+ \langle2nd\rangle}\ , \quad n=4\ .
\end{equation}
\item $V_{st}(V_{cb})$ also is proportional to the first order
perturbations but in the different form:
\begin{equation}
V_{st}\sim\langle1\rangle\ , \quad n=3\ ; \qquad V_{st}\sim \frac{
\langle2nd\rangle}{\langle1st\rangle}+\langle1st\rangle\ , \quad n=4\ .
\end{equation}
Because of such a difference $V_{us}$ turns out numerically larger than
$V_{st}$. The excessive factor contains the number of vector components
$-4$, besides unknown factors assumed to be of the unity order.

Thus, $|V_{us}|>|V_{st}|$ again might be an evidence for the vector
character of high energy interactions. But the difference is only
numerical here in contrast to usually  assumed differences in
perturbation orders \cite{10}.
\item
\begin{equation}
|V_{ub}|\ \le\ \langle3rd\rangle\ , \; \mbox{ the same with } |V_{td}|.
\end{equation}
Thus, the discussed mechanism reproduces observed differences in
hierarchy steps:
\end{enumerate}
\begin{equation}
\left|\frac{V_{us}}{V_{ud}}\right|\ \gg\ \left|\frac{V_{ub}}{V_{us}}
\right|\ .
\end{equation}
The interesting point here is that the necessary condition for such a
property seems to be the simultaneous presence of high energy
interactions dependent on flavors (perturbative, like responsible for
$\delta\beta_{ik}$ in Eq.(28)) and independent ones (may be strong,
responsible for the symmetric part $\lambda$ in Eq.(2)) [9(2)].

There are also some other properties and consequences of the mechanism.
\begin{itemize}
\item There are no direct relations between mass ratios and mixing
elements but there is correspondence of values with equal orders of
perturbations.
\item The $CP$ violation may naturally be included in the mixing matrix
$[9(2)]$. The ratios Im$V_{cd}/$Re$V_{cd}$, Im$V_{ts}/$ Re$V_{ts}$ can
be evaluated exclusively through mass ratios. Their numerical values
closely approach experimentally acceptable figures.

In establishing of this property a vector character of the interaction
distinguishing flavors is crucial again. The result is correct for any
$a_{nm}$ coefficients.

\item The interesting point is that the 4th generation of quarks (if it
exists)  insignificantly influences the properties of the three
generations. 

In addition the solution of the gap-\-equation with the hierarchy
spectrum permits also to obtain some parametric suppression of flavor 
changing neutral currents ($ \sim N^{-1}_c f_1(m_2/m_1)^{1/2}$, where 
$f_1$ is the function from Eq.(18)). Besides, the Goldstone bosons (18) 
necessary contain the highest flavor.  
\end{itemize}

All these facts permit us to state our final conclusion: the observed
form of the $V_{CKM}$ mixing matrix may evidence that quark generations
are distinguished due to new chiral vector interactions (neutral in
flavors), very likely without the scalar Yukawa couplings. The new
interaction could be gauge one, its scale $M$ is higher or much higher
than 1 TeV.

\newpage

\noindent
FIGURE CAPTIONS

\begin{description}
\item[Fig.1.] The gap equation in the leading $N_c$ approximation.
\item[Fig.2.] The self-energy and gap equations in the next after
leading $N_c$ approximation.
\item[Fig.3.] The $q\bar q$ scattering amplitude in the leading $N_c$
approximation.
\item[Fig.4.] The diagram for $R-L$ interactions through vector
exchanges (perturbative $g_i(L)$, $\tilde g_k(R)$ and unperturbative
$g$). The hatched area symbolizes the necessary strong coupling
independent of flavors.
\end{description}

\begin{figure}[t]
\epsfysize=200mm
\centerline{\epsfbox{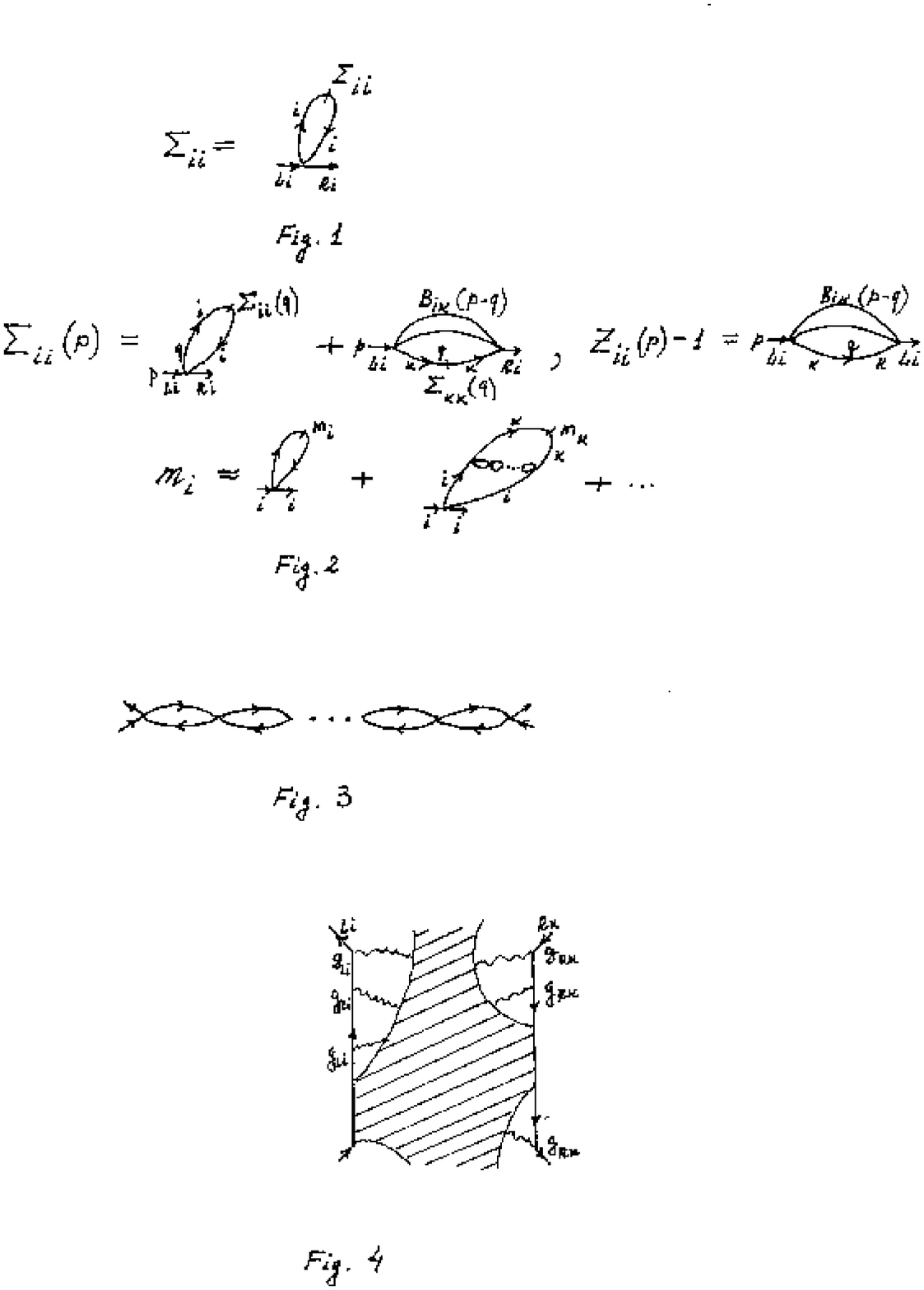}}
\end{figure}

\end{document}